\documentclass[article]{jss}
\usepackage[utf8]{inputenc}

\author{
Patrick Ng\\ \And Drew Plant\\ \And Yiran Sheng\\
}
\title{Do Price Ranges Increase Click-throughs?}

\Abstract{
An important goal of online comparison shopping services is to
``convert'' a viewer from general product category pages (for example
product groups such as ``smartphones'' or ``air-conditioners'') to
detailed product pages and ultimately to order pages. Comparison
shopping websites provide a familiar web interface as well as a chance
for consumers to purchase items at competitive prices. In return for
providing access to a large market of potential consumers, the
comparison shopping service usually receives financial compensation for product clicks and orders.  This study looked at 2.5 million product listing visits at price.com.hk
to determine whether a modification in the way prices are displayed on
general category pages resulted in more ``conversions'' to product
detail pages. We found a statistically significant improvement over-all
as a result of the new price display resulting in 3.6\% more product
clicks over all categories. Additional analysis showed that the effect
is heterogeneous among different categories, and in a few cases there
may be some categories negatively affected by the display modification.
}

\Plainauthor{Patrick Ng, Drew Plant, Yiran Sheng}
\Plaintitle{A Capitalized Title: Something about a Package foo}
\Shorttitle{Do Price Ranges Increase Click-throughs?}

\Submitdate{}

\Address{
      }

\usepackage{amsmath} \usepackage{graphicx} \usepackage{chngcntr}
\usepackage{scrextend} \usepackage[utf8]{inputenc}
\counterwithin{figure}{section} \counterwithin{equation}{section}
\counterwithin{table}{section} 

\begin{document}

\section{Introduction}\label{introduction}

Comparison shopping websites have the potential for providing users the
convenience of easily searching for competitive prices and evaluating
different related products in a very efficient manner. Among comparison
shopping websites there is competition for viewers as highlighted by
Belleflamme.

Price.com.hk is a top ecommerce website in Hong Kong where users can
find a variety of products and compare prices offered by different
sellers. These products are from a wide range of categories, including
mobile phones, digital cameras, TV's, washers, heaters, air
conditioners, diapers, baby formula, computers, and toys.

\begin{figure}[!ht]
  \centering
  \includegraphics[width=8cm]{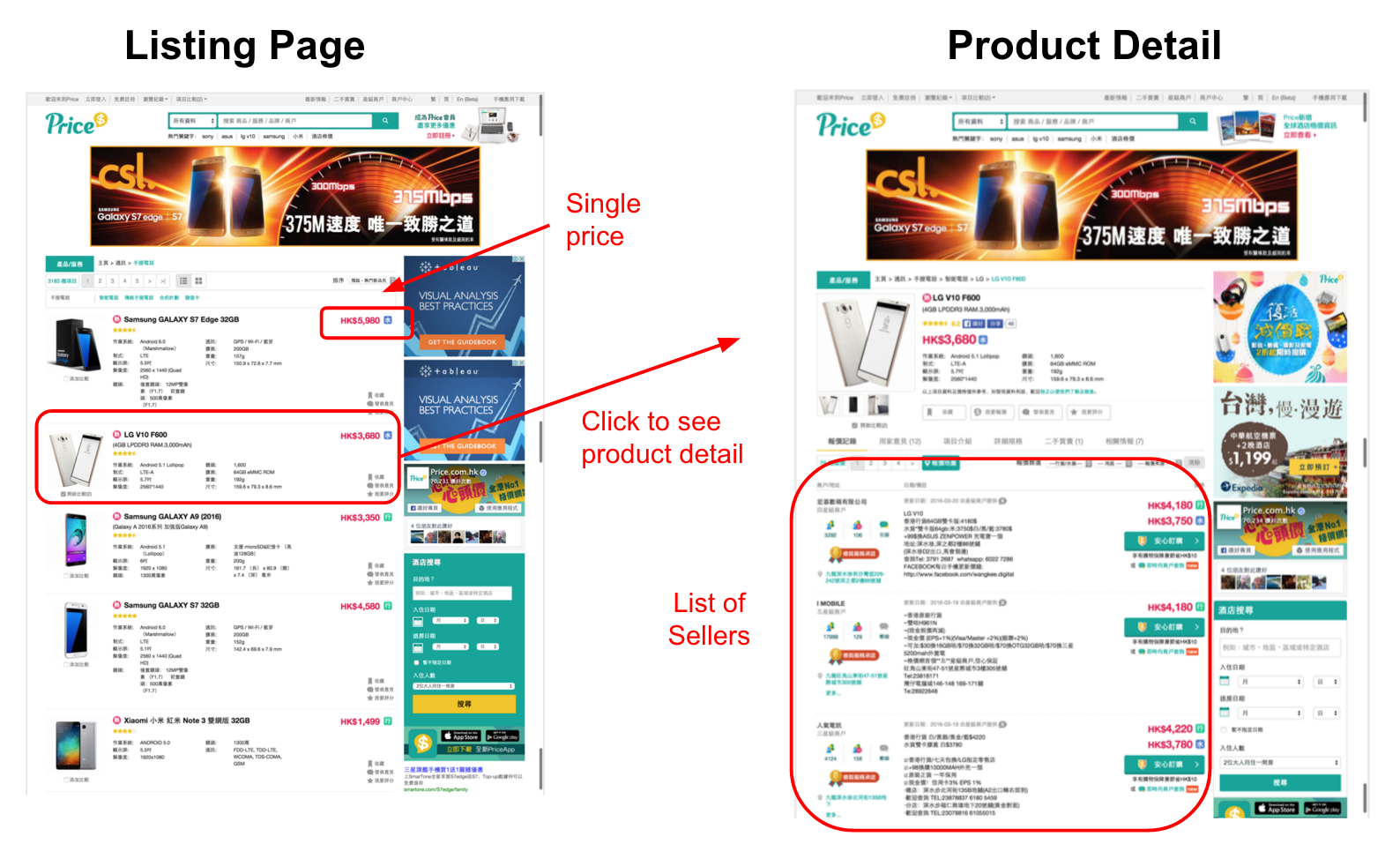}
  \caption{a category listing page click opens up a product detail page}
  \label{fig:ui}
\end{figure}

A typical user of Price.com.hk is interested in only one or a few
product categories. On a category listing page, a user can see a list of
products, together with a singular price for each product. When a user
clicks on a product, she sees a detailed product page including more
product information along with a list of all sellers and offered prices.
An example category listing page and resultant ``click-through'' product
detail page are shown in Figure \ref{fig:ui}. For each product in the
listing page, the single price shown represents the latest updated price
from a pool of sellers.

An inexperienced user may not realize that the single price shown
represents only a sample from potentially many competing sellers. What
if we display the price range from all sellers instead? Will that
attract more users to click on the product? Our research strives to
answer this question through a controlled field experiment. Several
studies have been carried out exploring the effects of displaying
product prices on web-sites towards purchasing behavior. For example,
Dholakia \& Simonson showed that in online auction sites, flanking a
product and its bid-price with comparison prices for the same product
tends to move the average bid-price for the product in the direction of
the flanking (comparison) prices. However in our experience this is the
first attempt to study how moving from a single-price display to price
ranges for each product on a product category page impacts user
click-throughs to detailed product pages.

There are a few reasons to suspect that displaying a price range rather
than a single price for products on a category listing page might result
in more product page clicks:

\begin{enumerate}
\def\labelenumi{\arabic{enumi}.}
\item
  Having a range of prices for each product enables users to get a
  better feel for how their product of interest stacks up against other
  products. For instance if I'm keen on a particular camera and I see
  that its price overlaps with that of another camera of obviously
  inferior quality and status, I might likely be more inclined to
  investigate my camera further, knowing that some retailers are
  offering it at a competitive price.
\item
  On the other hand if I see only single price for my ``dream'' camera
  and the inferior one I may choose alternate means (in the form of
  online comparison searches at alternative site), especially if I
  believe that only one price for my favorite camera is being offered
  through the current comparison shopping site.
\end{enumerate}

In the same vein, it is reasonable to hypothesize that different product
categories might show different sensitivity to this type of price-range
``treatment''. In particular, online shoppers are likely to spend more
time in detailed comparisons of big-ticket items such as TV's or
smart-phones than small-price consumables such as diapers in order to
eke out maximum feature satisfaction at the best price.

Our research studies data from 2.5 millions listing page views,
collected from a 20-day period in April 2016. Given the large sample
size, our experiment was able to resolve treatment effects that are both
statistically significant and practically important.

\section{Experimental Design}\label{experimental-design}

This experiment randomly assigned a different method for displaying
product prices on an aggregated product category listing page. By
design, treatment assignment was done at the user level. Here is an
explanation of the original price display method vs.~treatment price display method
along with our measurement outcome:

\hfill

\begin{addmargin}[3em]{1em}

\noindent \underline{Original method}: A \textbf{single} price is displayed for each product position on the category page.  This single price represents the last modified price offered by any retailer for this product.  There are a maximum of 15 products displayed on a listing page.  Here is an example of the original price for a particular-model of smart phone:

\begin{figure}[!ht]
  \centering
  \includegraphics[width=2.8cm]{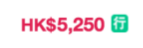}
  \caption{Single price UI}
  \label{fig:single}
\end{figure}

\noindent \underline{Treatment method}: Each product displays a price \textbf{range} (low-price - high-price) range of prices currently offered for each respective product.  In fact two ranges are listed: one official  product range and one so-called ``water-price"\footnote{``Water-price" refers to the price for parallel import items, which typically don't carry warranties but are usually offered for cheaper prices than corresponding official market items.} range.  Using our same smart-phone example the price display applying our treatment looks like:

\begin{figure}[!ht]
  \centering
  \includegraphics[width=4cm]{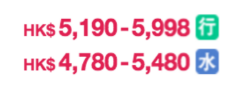}
  \caption{Price range UI}
  \label{fig:range}
\end{figure}

Figure \ref{fig:left-right-ui} compares category listing pages using both the control and treatment price display methods.

\begin{figure}[!ht] 
  \centering
  \includegraphics{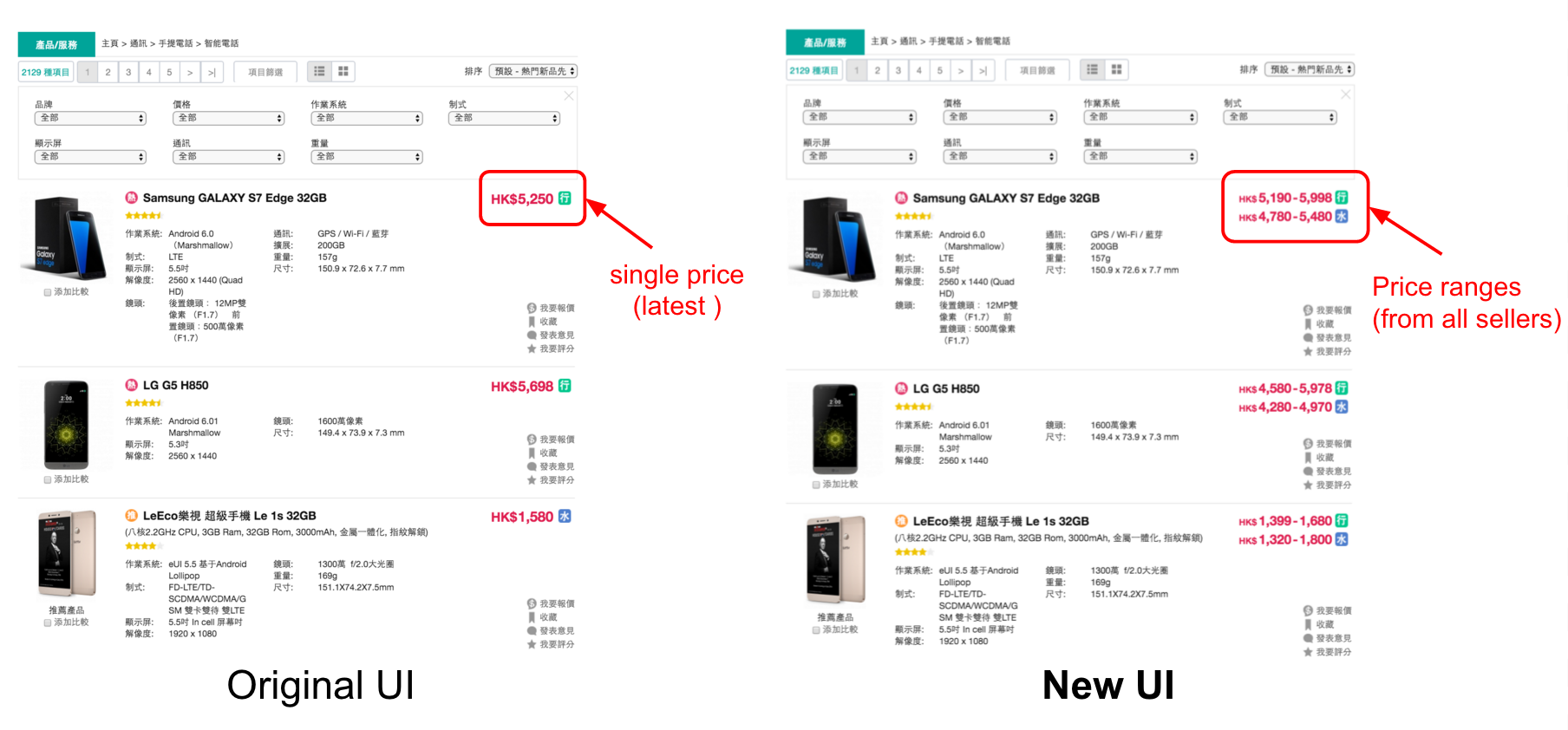}
  \caption{The control (left) and the treatment (right) prices on category listing page}
  \label{fig:left-right-ui}
\end{figure}

\noindent \underline{Outcome measure}: We measured the number of clicks (those which open detailed individual-product pages) for each category listing page and each user.  Observations were made at the level of unique user and individual category page for that user.  Therefore, if Jane Chan clicks to three detailed product pages from her first listing page of  category “smart-phone”, she scores a \emph{clicks = 3} for that category listing page.

\end{addmargin}

\hfill 

In price.com.hk parlance, the term \emph{category} refers to the first-level
aggregation of product types. The aforementioned price-range treatment
was applied to category product listing pages randomly assigned to
receive treatment. In fact, price.com.hk has a product classification
hierarchy with headings of zone → group → category → product. Simply
put, our experiment applied treatment at the category level and involved
subsetting of data at the zone level as will be discussed later.

Our web-logs provided a rich set of other attributes such as
\texttt{productPos}, a product's position on a category page (typically
1-15) and \texttt{pageNumber} within a particular \emph{category}
listing. Also included are (1) status on whether a user is logged in and
(2) the browser being used. These other attributes provide rich food for
further consideration towards discovering other potentially interesting
treatment interactions.

As previously mentioned there is reason to believe that the treatment
might result in less user confusion and greater potential for the user
to click-advance to one or more product detail, thus resulting in more
order-potential for products, more ad exposure in general and more
revenue for the comparison shopping service provider.

Our data was collected in the time period in 2016 from Apr 1 - Apr 25.
The data consisted of 6 million records from web-logs during that period.
Our randomized treatment was applied on Apr 6, 2016, so entry rows prior
to that period show the old single-price range display. Treatment was
randomized at the level of a user's rendered product category listing
page (represented by the variable sessionId)\footnote{Initially, the
  design specified that treatment be assigned through a cookie
  randomized at the user-browser level. Unfortunately cookies were
  erroneously re-assigned at periodic intervals, causing some users to
  have multiple assignments. Nonetheless assignments remained unique at
  the listing session level. As discussed in Section \ref{complication}
  we don't believe this anomaly affected the essential part of our
  conclusions.}.

Table \ref{tab:summ} provides a top-level summary of our experiment:

\begin{table}[!htbp]
\centering
\caption{Experiment Summary}
\label{tab:summ}
\begin{tabular}{@{\extracolsep{5pt}} l|r}
\hline\hline\\[-1.8ex]
                                            & \textbf{The Experiment} \\ \\[-1.8ex]\hline\\[-0.2ex]
{Time period:  }                         & April 6-25, 2016     \\ 
{Length of Experiment:  }                        & 20 days        \\ 
{Number of category views observed:  }          & 2,482,654      \\ 
{Number of product click-throughs observed:  }             & 773,450        \\ 
{Mean product click-throughs for control group:  }   & 0.305          \\ 
{Mean product click-throughs for treatment group:  } & 0.316          \\ [1.8ex]\hline\hline
\end{tabular}
\end{table}

To test that randomization was done properly, A/A testing was carried
out by modeling treatment assignment against category grouped at the
level of user (\texttt{anonyId}) using data prior to the actual
treatment application (in other words, prior to the cut-over to price
range on Apr 6.) As discussed in Section \ref{models}, A/A stage placebo
test indicated that randomization was valid over a variety of observed
potential exogenous covariates.

For A/B analysis, we studied the effect of price range display on number
of clicks per category listing page. This was first studied in a simple
model which regressed clicks against the single binary variable treat
(where 0 == ``control'' and 1 == ``treatment''.) We then used a more
involved model which also included as a covariate the average click-rate
from each category listing page without treatment (from A/A data). In
general we expected that different categories have different visit
average rates and so including this base level rate as a function of
different categories might reduce our model overall standard error. We
then had a third set of A/B models which involved subsetting by zone
which is the largest group level of products. As an example, the
\textbf{Communication} zone includes \textbf{smartphone} category along
with other categories Service Plan and Prepaid Sim. Subsetting our
models by one zone at a time to consider only categories within the
respective zone greatly simplified our model computation and allowed us
to focus on those zones showing the most significant effects.

Power analysis was done in order to estimate number of samples required
for treatment effects of interest. Based on our A/A stage base-level
average outcome rate of 0.32 clicks per listing session, we were
interested in resolving treatment effects on the order of 3\%. Our power
analysis indicated that with a minimum of half a million samples we
could resolve a treatment effect of 3\% ($\alpha = 0.05$) with a
statistical power of 99\%. Given that our record set has over 2 million
listing sessions (after cleanup), we felt that we had a good
experimental setup for resolving an effect at at least 3\% of our
baseline.

\section{Data Collection}\label{data-collection}

We primarily rely on server side tracking technologies to gather
experimental data. Raw data is collected through web server logs. For an
overview of web tracking technologies and their usage in the industry,
Schmucker (2011) provides a detailed summary. We will outline relevant
techniques employed in our study in this section.

Server side tracking is usually conducted by including a small payload
for tracking purposes alongside normal application code. Payloads are
typically injected in two places:

\begin{enumerate}
\def\labelenumi{\arabic{enumi}.}
\itemsep1pt\parskip0pt\parsep0pt
\item
  Page url query strings
\item
  Browser cookies
\end{enumerate}

A ``query string'' is part of a URL and can be recognized by the fact
that it is not part of the normal URL hierarchical path. It contains an
arbitrary number of named parameters along with their values. The query
string is processed by the server application to locate specific
resources requested by the client in conjunction with the rest of the
URL. Within the query string an owner of a website can append additional
parameters for purposes of tracking or providing context to a page
request. These added parameters will be ignored by web servers and not
affect normal site operation; however, they are faithfully recorded by
server logging for retrieval and analysis later. As an example consider
the following hypothetical URL query string:

\begin{verbatim}
?product_id=123&click_type=thumbnail
\end{verbatim}

This query string contains two parameters: \texttt{product\_id} and
\texttt{click\_type}. \texttt{product\_id} is a request for the web
server to retrieve the specific product page a user is requesting.
\texttt{click\_type} is an added parameter for usage tracking, and has
no user-discernible effect on the web page returned. Rather, it is used
to indicate the type of link which was used to open up the current page;
in our example the user clicked a thumbnail image to navigate to the
product page. This tracking parameter thus provides rich additional
context to the page request.

An HTTP ``cookie'' is a small piece of data sent from a website and
stored in the user's web browser while the user is browsing. Every time
the user loads the website, the browser sends the cookie back to the
server thus notifying the server of the user's activity history. Similar
to query strings, cookies can be used both to serve functional and
tracking purposes. For instance, authentication cookies keep track of
whether a user is logged in or not, and if so which account she has been
using. Tracking cookies, on the other hand, are used to build a
long-term history of a given user's browsing behavior. In its most basic
form, a tracking cookie may assign a random but persistent unique id to
a user when she first visits a website. Later when she returns the
cookie is sent back to the server, signaling a page is requested by a
returning visitor.

In our experiment, each user is assigned two persistent cookies:
\texttt{anonyId} and \texttt{treat}. \texttt{anonyId} is a unique
identifier for each visitors and \texttt{treat} is a binary indicator of
whether this visitor is assigned to control or treatment group. After a
user's first visit to price.com.hk, subsequent page requests are
accompanied by these two pieces of data which are sent to the server and
recorded in web logs. The server examines the \texttt{treat} cookie and
(based on the value of treat) creates listing pages with or without
treatment UI. In theory, this setup assigns treatment at the visitor
level. A new visitor is randomly assigned to control or treatment, and
her treatment remains the same for all subsequent visits to the site.

In addition to managing treatment assignment, we need to label each
product page with the category listing from which that product page was
requested. This ``parent-page'' information is tracked using query
strings. When a visitor requests a category listing page, the server
creates a unique session id, and appends the id inside of a query string
to all product URL links on the listing page. This \texttt{sessionId}
parameter is applied in query strings for both control and treatment
groups.

One interesting data-logging challenge was how to associate
\texttt{sessionId} with the category page it identifies. The parameter
\texttt{sessionId} is unique to each category listing visit; however it
is generated \emph{after} a user requests that listing page. Therefore,
when the server first generates a listing page and and enters a
corresponding web-log there is no sessionId available to attach to the
log entry.. To retroactively assign \texttt{sessionId} to a listing
page, a special technique is used. On every listing page an invisible,
one-pixel image is added at the bottom. The url of the one-pixel image
contains a query string which includes the generated \texttt{sessionId}.
The browser makes a request for that image resulting in
\texttt{sessionId} being passed as a query string parameter and thereby
getting logged alongside the initial listing-page session web-log
entry\footnote{The ``tracking pixel'' is appended to the bottom of a
  page so as to not degrade overall site performance. However,
  asynchronous image loading (commonly done by most browsers) can result
  in the sessionId not getting logged. This will happen for example if a
  user loads a listing session page and immediately clicks on a product
  link before the tracking pixel can be loaded by the browser. This
  would result in an orphaned product click (with no identifying
  \texttt{sessionId}). We removed occurrences of these orphaned product
  clicks from our data prior to analysis.}. This technique is commonly
employed by web tracking tools such as Google Analytics.

\begin{figure}[!ht]
  \centering
  \includegraphics{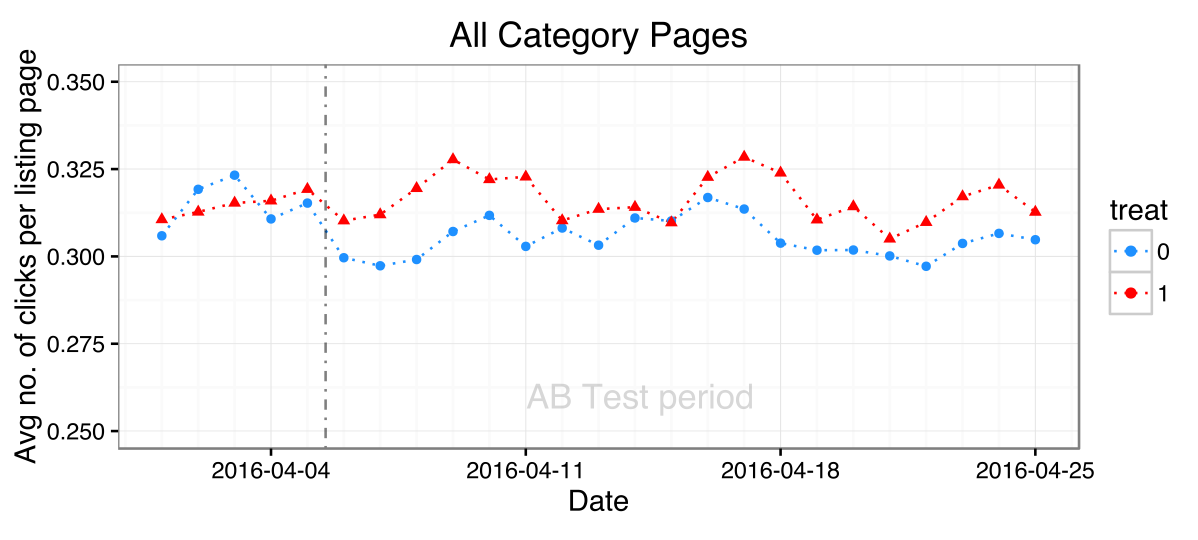}
  \caption{Daily Click-through for Control and Treatment Groups}
  \label{fig:dailyclicks}
\end{figure}

Figure \ref{fig:dailyclicks} provides an overview of click-through data
over the entire experiment period for control and treatment groups.

\subsection{Complications on Data
Integrity}\label{complications-on-data-integrity}

\label{complication}

It's not uncommon when analyzing web traffic data to encounter
complications with data hygiene and implementation / clerical errors.
Our experiment was no exception.

\begin{figure}[!ht]
  \centering
  \includegraphics[width=8cm]{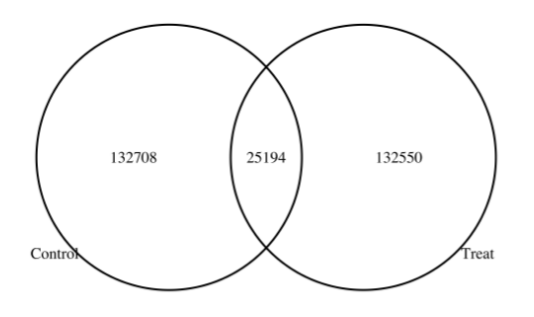}
  \caption{Venn diagram of anonyId counts}
  \label{fig:venn}
\end{figure}

As mentioned previously, during the A/A data analysis stage, we
discovered that a non-trivial portion of visitors were mistakenly
assigned to both treatment and control group. We reckoned this was a
consequence of users clearing cookies or some unknowable technical
factors. Upon further discussion with the engineering team, we confirmed
this was due to an implementation bug: cookies had been set to expire
resulting in treatments being reassigned every 90 hours. Figure
\ref{fig:venn} illustrates the scale of this issue.

Users with re-assigned treatment accounted for 22\% of total traffic.
Moreover 38\% of listing sessions were generated by users having
treatment status which varied during the course of the experiment. This
issue raised a new challenge to our analysis. We reasoned that this does
not add significant bias to our analysis when estimating ATE, either
intuitively or statistically. Our rationale is that treatment should
only affect the outcome for the current listing page (not future listing
pages which may have a different treatment assignment). In other words
we believe that there should be no cross-session spillover effects. For
a more in-depth discussion, refer to Appendix \ref{reassign}.

Another challenge to maintaining data integrity was garbage data
generated by bot traffic. Many automated systems such as web crawlers
generate a large volume of web traffic, and we are not interested in
treatment impact on their behavior. We removed all known bots (ie. those
self-declared through user agent in web logs) and utilized a few
conservative heuristics in detecting non-obvious bots. For instance we
noticed that a very small number of users logged an abnormally large
number of web requests and displayed unusual click patterns, for
instance showing an equal number of clicks for every product in each of
the 15 page positions for every page of the listing category.
Unfortunately this approach may result in a few false negatives since
sophisticated bots which are trained to spoof their user agent can
easily bypass such heuristics.

All these complications may impact results in subtle ways. While we did
our best to mitigate discovered issues, it's worth keeping them in mind
when interpreting and generalizing results.

\section{Models}\label{models}

\label{models}

Prior to implementation of the experiment we carefully considered
choices for how best to model the behavior of interest. Three key
considerations to achieving this goal were:

\begin{itemize}
\itemsep1pt\parskip0pt\parsep0pt
\item
  how best to ensure randomization was done correctly,
\item
  how to model our outcome measure so as to provide a regression model
  which is defensible as most closely representing reality, and
\item
  how to consider and handle error heteroskedasticity.
\end{itemize}

\subsection{Randomization Validation}\label{randomization-validation}

Since random assignment was performed at user/browser level, we wanted
to validate there were no systematic differences in browser-level
measures between control and treatment group. Randomized assignment was
done at the A/A stage of data collection (before actual treatment was
done). We used A/A stage data therefore to perform a placebo test. A
logistic regression model was used to study whether browser-level
measures correlate with treatment assignment. See equation
(\ref{eq:model-valid}).

\begin{equation} \label{eq:model-valid}
\text{treat} \sim 1 + \text{UA} +  \text{categoryCountPerAnonyId} 
\end{equation}

Model (\ref{eq:model-valid}) regresses treat against
categoryCountPerAnonyId which counts the number of category pages
accessed by user-browser. This regression produced 252 coefficients (one
for each category) along with p-values indicating how significant each
category was towards increasing or decreasing likelihood of treatment
assignment. Noticeably, only 12 of these categories (4.76\%) are
significant below the $\lambda = 0.05$ level. However, due to chance
alone and by definition of type I errors, we expect 5\% of them to be
accidentally significant. Using the Bonferroni correction (requiring a
p-value of 0.05 / 252 = 0.000198 for rejection of randomization defects)
our placebo test indicates that we don't need to reject the null
hypothesis. We therefore have no reason to suspect that randomization
was done improperly.

\subsection{OLS vs Poisson Regression}\label{ols-vs-poisson-regression}

OLS (Ordinary Least Squares) regression has proven to be a robust tool
in experimental studies due to its simplicity and flexibility. However,
in our experiment, some of the standard conditions for application of
OLS are violated to some degree. It was therefore necessary for us to
address how robust OLS models are against these deviations from standard
practice. One particular concern surrounded the normality assumption for
regression model residuals.

Clearly, the outcome measure in our experiment is \emph{not} normally
distributed. Observed click counts per session are discrete,
non-negative values, and have a long tail in their distribution. A good
candidate for modeling discrete non-negative random variables is the
Poisson distribution. We evaluated GLM (generalized linear model) using
Poisson covariance matrix (Poisson Regression) as an alternative to OLS
prior to A/B stage data collection. The downsides to using the Poisson
Regression are:

\begin{enumerate}
\def\labelenumi{\arabic{enumi}.}
\itemsep1pt\parskip0pt\parsep0pt
\item
  Use of Poisson residuals requires the assumption that outcome mean and
  variance are equal ($\sigma^2 = \mu$).\\
\item
  Poisson Regression parameter coefficients are hard to interpret when
  covariates are included due to nonlinear transformation of regression
  outputs.
\end{enumerate}

To weigh the relative merits of these two modeling approaches (OLS
vs.~Poisson), we conducted a statistical simulation. We generated random
data modeled from a Poisson distribution based on assumed treatment
effect, which manifests itself in the differences between the Poisson
distribution parameters $\lambda$ (average clicks per listing session)
of the control and treatment groups. Details of the simulation can be
found in the Appendix \ref{ols-sim}. The simulation showed that as the
true treatment effect used in generating simulated data increases, OLS
tends to underestimate the standard error of the treatment effect. On
the other hand, OLS produces nearly identical estimates of treatment
effect compared to Poisson Distribution\footnote{By Gauss--Markov
  theorem, OLS estimates are best linear unbiased estimators (BLUE).}.
In our experiment, we expected the treatment effect to be small compared
to the average level of outcome observed in A/A stage (0.33 clicks per
listing session), therefore we decided that underestimation of standard
error would probably not be an issue.

\subsection{Regression Models and Residual Covariance
Structure}\label{regression-models-and-residual-covariance-structure}

After settling on OLS as our main analytical approach, the base-line
model is given by (expressed in \proglang{R} notation):

\begin{equation} \label{eq:simple-model}
\text{clicks} \sim 1 + \text{treat}
\end{equation}

This model is equivalent to a two-sampled T-test with population
variance assumed to be equal between control and treatment group. This
assumption also is not entirely accurate. Note in our experiment, each
unit of observation/data point is a category listing page session,
rather than an individual typically seen in socio-economic studies. In
addition, our experiment is conducted at the category listing session
level, clustering on each user (identified by cookie \texttt{anonyId}).
Therefore, we need to correct for the fact that listing sessions which
are generated by the same user are correlated. We accounted for this by
applying correction clustering by \texttt{anonyId} to achieve robust
standard error estimates for the treatment effect(s). Throughout this
section, all models discussed have this clustering standard error
correction applied to them.

Our second model included the following additional covariates:

\begin{itemize}
\itemsep1pt\parskip0pt\parsep0pt
\item
  \texttt{catClickRateAA} (the average number of clicks per listing
  session per user computed from A/A stage data for this particular
  category)
\item
  \texttt{pageNo} (the page number of the category listing being viewed)
\item
  \texttt{itemsPerPage} (the number of products on the current category
  listing page)
\item
  \texttt{isLoggedIn} (whether the current user has logged into the
  website.)
\end{itemize}

It's worth noting that while we believe these covariates are useful for
explaining outcome clicks, they are not necessarily causal in their
contribution to model (\ref{eq:simple-model}). Among them,
\texttt{catClickRateAA} is worth special mention. This variable
represents average A/A stage data. Our hope was that a lot of variation
in click rate between categories could be explained by variation in
base-level click rates. Therefore,by including it as a covariate, we
hoped the overall regression model's explanatory power could be
improved, and standard error of the treatment effect coefficient could
be reduced.

\begin{equation} \label{eq:model-covariate}
\text{clicks} \sim 1 + \text{treat} + \text{pageNo} + \text{itemsPerPage} + \text{isLogin} + \text{catClickRateAA}
\end{equation}

Furthermore, the potentially heterogeneous effect of treatment on clicks
averaged at the zone level was deemed to have a particularly interesting
business interpretation. In other words, what is the specific impact of
displaying a price range on average clicks within each large grouping of
products? We therefore estimated an individual model
(\ref{eq:model-covariate}) using only data for each zone. 

In fact we could have combined heterogeneous treatment effects for all
zones into a unified model and estimated it on the entire dataset.
However, in doing so, we would have needed to force the assumption that
residual terms have identical variances for all zones, which we believe
is unreasonable. See Appendix \ref{zone} for further discussion.

\pagebreak
\section{Treatment Effect}\label{treatment-effect}

A comparison of model (\ref{eq:simple-model}) and model
(\ref{eq:model-covariate}) is shown in Table \ref{tab:reg-main}.

\begin{table}[!htbp] \centering 
  \caption{Effect of display of Price Range on Product Click-through} 
  \label{tab:reg-main}  
\begin{tabular}{@{\extracolsep{5pt}}lcc} 
\\[-1.8ex]\hline 
\hline \\[-1.8ex] 
 & \multicolumn{2}{c}{\textit{Dependent variable:}} \\ 
\cline{2-3} 
\\[-1.8ex] & \multicolumn{2}{c}{Product Click-through} \\ 
 & Basic & Full \\ 
\hline \\[-1.8ex] 
 treat & 0.011$^{***}$ (0.002) & 0.011$^{***}$ (0.002) \\ 
  pageNo &  & $-$0.008$^{***}$ (0.001) \\ 
  itemsPerPage &  & $-$0.004$^{***}$ (0.0002) \\ 
  isLogin &  & 0.076$^{***}$ (0.007) \\ 
  catClickRateAA &  & 0.870$^{***}$ (0.009) \\ 
  Constant & 0.305$^{***}$ (0.001) & 0.105$^{***}$ (0.004) \\ 
 \hline \\[-1.8ex] 
Observations & 2,482,654 & 2,438,223 \\ 
R$^{2}$ & 0.0001 & 0.016 \\ 
Adjusted R$^{2}$ & 0.0001 & 0.016 \\ 
Residual Std. Error & 0.766 (df = 2482652) & 0.765 (df = 2438217) \\ 
F Statistic & 135.000$^{***}$ (df = 1; 2482652) & 8,124.000$^{***}$ (df = 5; 2438217) \\ 
\hline 
\hline \\[-1.8ex] 
\textit{Note:}  & \multicolumn{2}{r}{$^{*}$p$<$0.1; $^{**}$p$<$0.05; $^{***}$p$<$0.01} \\ 
 & \multicolumn{2}{r}{SE are cluster-robust, clustering on per-browser-instance} \\ 
\end{tabular} 
\end{table}

Both models yielded similar treatment effect estimates of 0.011. On
average, the treatment (price range) boosts clicks by 0.011 which
amounts to a 3.6\% increase. Adding covariates to the simple model
(\ref{eq:simple-model}) increased overall explanatory power of the
regression (evidenced by the apparent improvement in $R^2$); however
contrary to our expectation, the standard error of treatment effect
remains the same, after adjusting for robust clustered standard errors.

It's worth considering the practical ramifications of these results: The
estimated treatment effect translates into an expected increase of 1,268
daily product page views, with a 95\% confidence bound of (807,
1,729)\footnote{Based on A/A stage daily average of listing session
  counts across all categories (averaging 115,285.5).}. To put these
numbers into perspective, the normal fluctuation of daily listing
session count is around +/-7,000, as measured by standard deviation.
Zone-level treatment effects are reported in Table \ref{tab:reg-zone23}
and Table \ref{tab:reg-zone46}, showing the four zones with highest
traffic volume.

\begin{table}[!htbp] \centering 
  \caption{Effect of display of Price Range on Product Click-through (Zone 2 \& 3)} 
  \label{tab:reg-zone23} 
\begin{tabular}{@{\extracolsep{5pt}}lcc} 
\\[-1.8ex]\hline 
\hline \\[-1.8ex] 
 & \multicolumn{2}{c}{\textit{Dependent variable:}} \\ 
\cline{2-3} 
\\[-1.8ex] & \multicolumn{2}{c}{clicks} \\ 
 & Zone 2 - Camera and Photo & Zone 3 - AVl \\ 
\hline \\[-1.8ex] 
 treat & $-$0.012$^{***}$ (0.003) & 0.011$^{***}$ (0.003) \\ 
  pageNo & $-$0.012$^{***}$ (0.0004) & $-$0.006$^{***}$ (0.0003) \\ 
  itemsPerPage & 0.00000 (0.0005) & $-$0.001$^{***}$ (0.0004) \\ 
  isLogin & 0.057$^{***}$ (0.007) & 0.075$^{***}$ (0.008) \\ 
  catClickRateAA & 0.859$^{***}$ (0.017) & 0.858$^{***}$ (0.014) \\ 
  Constant & 0.062$^{***}$ (0.009) & 0.066$^{***}$ (0.007) \\ 
 \hline \\[-1.8ex] 
Observations & 308,331 & 255,811 \\ 
R$^{2}$ & 0.012 & 0.018 \\ 
Adjusted R$^{2}$ & 0.012 & 0.018 \\ 
Residual Std. Error & 0.745 (df = 308325) & 0.737 (df = 255805) \\ 
F Statistic & 745.000$^{***}$ (df = 5; 308325) & 937.000$^{***}$ (df = 5; 255805) \\ 
\hline 
\hline \\[-1.8ex] 
\textit{Note:}  & \multicolumn{2}{r}{$^{*}$p$<$0.1; $^{**}$p$<$0.05; $^{***}$p$<$0.01} \\ 
 & \multicolumn{2}{r}{SE are cluster-robust, clustering on per-browser-instance} \\ 
\end{tabular} 
\end{table}

\begin{table}[!htbp] \centering 
  \caption{Effect of display of Price Range on Product Click-through (Zone 4 \& 6)} 
  \label{tab:reg-zone46} 
\begin{tabular}{@{\extracolsep{5pt}}lcc} 
\\[-1.8ex]\hline 
\hline \\[-1.8ex] 
 & \multicolumn{2}{c}{\textit{Dependent variable:}} \\ 
\cline{2-3} 
\\[-1.8ex] & \multicolumn{2}{c}{clicks} \\ 
 & Zone 4 - Computer & Zone 6 - Home Appliance \\ 
\hline \\[-1.8ex] 
 treat & 0.012$^{***}$ (0.001) & 0.029$^{***}$ (0.002) \\ 
  pageNo & $-$0.007$^{***}$ (0.0002) & $-$0.013$^{***}$ (0.0003) \\ 
  itemsPerPage & 0.003$^{***}$ (0.0002) & $-$0.013$^{***}$ (0.0003) \\ 
  isLogin & 0.063$^{***}$ (0.004) & 0.099$^{***}$ (0.006) \\ 
  catClickRateAA & 0.964$^{***}$ (0.010) & 0.765$^{***}$ (0.011) \\ 
  Constant & $-$0.023$^{***}$ (0.005) & 0.272$^{***}$ (0.006) \\ 
 \hline \\[-1.8ex] 
Observations & 958,688 & 615,977 \\ 
R$^{2}$ & 0.012 & 0.017 \\ 
Adjusted R$^{2}$ & 0.012 & 0.017 \\ 
Residual Std. Error & 0.707 (df = 958682) & 0.895 (df = 615971) \\ 
F Statistic & 2,251.000$^{***}$ (df = 5; 958682) & 2,132.000$^{***}$ (df = 5; 615971) \\ 
\hline 
\hline \\[-1.8ex] 
\textit{Note:}  & \multicolumn{2}{r}{$^{*}$p$<$0.1; $^{**}$p$<$0.05; $^{***}$p$<$0.01} \\ 
 & \multicolumn{2}{r}{SE are cluster-robust, clustering on per-browser-instance} \\ 
\end{tabular} 
\end{table}

The following interesting observations can be made from Table
\ref{tab:reg-zone23} and \ref{tab:reg-zone46}:

\begin{enumerate}
\def\labelenumi{\arabic{enumi}.}
\itemsep1pt\parskip0pt\parsep0pt
\item
  The treatment effect for each zone (Camera \& Photo, Audio-visual,
  Computer, and Home Appliance) is significant to the p \textless{} 1\%
  level.
\item
  The Camera\&Photo zone shows a negative ATE, indicating a potential
  reduction in average clicks from treatment.
\item
  Other zones show significant positive ATE's, suggesting improvement of
  click rate with treatment.
\end{enumerate}

\pagebreak
\section{Conclusion}\label{conclusion}

From this controlled experiment, we found that the UI change in the
product listing page increased the per-listing product click-through
from 0.305 to 0.316, an increase of 0.11 (3.6\%.) The effect is
heterogeneous across different product groupings or zones. Positive effects were observed in zones Audio-Video (+0.011), Computer (+0.012) .and
Home Appliance (+0.029) whereas a negative effect was observed in the Camera \& Photo zone
(-0.012).

Prior to the experiment, Price.com.hk expected the UI change would have
a pronounced positive effect on behavior of inexperienced users. The
idea is this: Inexperienced users may believe the single price shown on
a listing page is the only price available. They therefore may not be
motivated to click-through to the product-detail page. On the other hand
experienced users recognize that the price is just one price among all
prices from all sellers and feel more confident in clicking to view
other prices on the product page. Therefore the price-range treatment
expectedly is more valuable for inexperienced users. Price.com.hk
speculates that a large portion of users in the Home Appliance zone are
inexperienced. Our experimental results support that theory.

One notable exception to the otherwise positive observed treatments is
the negative ATE in the Camera \& Photo zone. We need to dig further
into per-category effects to confirm and further explain this negative
effect. Based on later findings we might suggest modifying or perhaps
removing the price-range treatment for category pages in this zone.

Finally it is worth stressing that among the explanatory variables used
in the model, only \texttt{treat} has a causal interpretation on the
change in outcome clicks. Other model variables such as category level
average click through do not have a causal interpretation despite their
relatively large coefficients. These other variables are included in the
model as covariates to reduce the standard error of the \texttt{treat}
variable.

\section{Recommendations for Future
Experimentation}\label{recommendations-for-future-experimentation}

A common complaint regarding A/B testing for web-page changes on
conversion rate is the difficulty in drawing conclusions on persistence
of the treatment effect. Given the short duration of our study we were
not able to include measures of ATE persistence, though given a longer
study period this perhaps would be possible. Two competing hypotheses
could explain the increase in click-through rate: (1) our treatment led
to higher click-through which persisted over time -or- (2) the novelty
factor of treatment temporarily boosted click-through but the effect
eventually died out. A follow-up study conducted over a longer time
period of several weeks or months could better discriminate persistence
in long-term treatment effects.

Given that our experiment showed different ATEs for different zones,
it would be natural to prescribe treatment only for those zones showing positive ATEs but not for 
the zones showing negative ATE.
However, our experiment setting only varied treatment assignment at user
level without regard to zones. It's worth considering what a
zone-conditional application of treatment might mean on outcomes: if an
individual user sees price ranges on category pages within a ``treatment
zone'' but only single prices on category pages within a ``non-treatment
zone'' it might cause confusion and lower over click-through rates. The risk of ``cross-zone spill-over" 
is something to be considered prior to conditionally applying treatment,
though our current experimental setup cannot formally test for this
possibility.

To understand the effects of non-uniform treatment application, a
followup experiment could be carried out randomizing treatment across different zones and different users. In this
scenario, one group could have treatment applied across all zones
whereas another treatment group see treatment only for one of the eleven
zones. Such an experiment would potentially provide insight helpful in
deciding whether to introduce price range treatment only for selected
zones.

Finally it would be interesting to include the effects of price (and
price ranges) on click-through rate. In our experimental setup we only
had access to prices of products which were clicked but not for products
listed on the category page which were not clicked (ie. with a
click-rate of 0.) A future experiment whereby 0-click product pricing
were made available (ie by joining web log data with product data) could
answer interesting questions regarding treatment effects from different
price levels or price ranges.

\section{Appendix}\label{appendix}

\subsection{OLS vs Poisson Regression
Simulation}\label{ols-vs-poisson-regression-simulation}

\label{ols-sim}

Let the number of clicks per session exposed to
control be denoted by $\mathbf{Y}(0)$ and the number of clicks per session exposed to treatment by $\mathbf{Y}(1)$. Let
$\mathbf{D}$ be the randomized treatment assignment. We make the
following assumptions regarding population:

\begin{itemize}
\itemsep1pt\parskip0pt\parsep0pt
\item
  $\mathbf{D} \sim \text{Bernoulli(0.5)}$
\item
  $\mathbf{Y}(0) \sim \text{Poisson}(\lambda_0)$
\item
  $\mathbf{Y}(1) \sim \text{Poisson}(\lambda_1)$
\item
  $\lambda_1 = \lambda_0 + \text{ATE}$, $\text{ATE} \neq 0$
\item
  Denote observed click rate $\mathbf{Y}$, and
  $\mathbf{Y} = \mathbf{Y}(1) \cdot \mathbf{D} + \mathbf{Y}(0)(1-\mathbf{D})$
\end{itemize}

Operationally, we generate observations 5000 control observations
$y_i|d_i=0$ from Poisson distribution with $\lambda_0 = 0.35$. This
number is picked by computing A/A stage overall average click through.
In addition, 5000 hypothetical treatment observations $y_i|d_i=1$ are
generated with $\lambda_1 = \lambda_0 + \text{ATE}$, where ATE vary
between $0.001$ and $0.35$ (0.2\% to 100\% of $\lambda_0$). The mixed
$y_i$'s are regressed on $d_i$, with both OLS and Poisson Regression.

The OLS formulation of the regression model is given by:

\newcommand{\Expect}{{\rm I\kern-.3em E}}

\begin{equation} \label{eq:sim-ols}
\Expect[\mathbf{Y}|\mathbf{D}] = \alpha_{\text{OLS}} + \beta_{\text{OLS}} \mathbf{D}
\end{equation}

We have:

\begin{equation} \label{eq:ols-ate}
\hat{\text{ATE}} = \hat{\beta_{\text{OLS}}}
\end{equation}

And Poission Regression model:

\begin{equation} \label{eq:sim-glm}
\log(\Expect[\mathbf{Y}|\mathbf{D}]) = \alpha_{\text{POIS}} + \beta_{\text{POIS}} \mathbf{D}
\end{equation}

To derive estimate of $\hat{\text{ATE}}$ from Poisson Regression:

\begin{equation} \label{eq:pois-ate}
\hat{\text{ATE}} = \exp(\hat{\alpha_{\text{POIS}}} + \hat{\beta_{\text{POIS}}}) - \exp(\hat{\alpha_{\text{POIS}}})
\end{equation}

\begin{figure}[!ht] 
  \centering
  \includegraphics{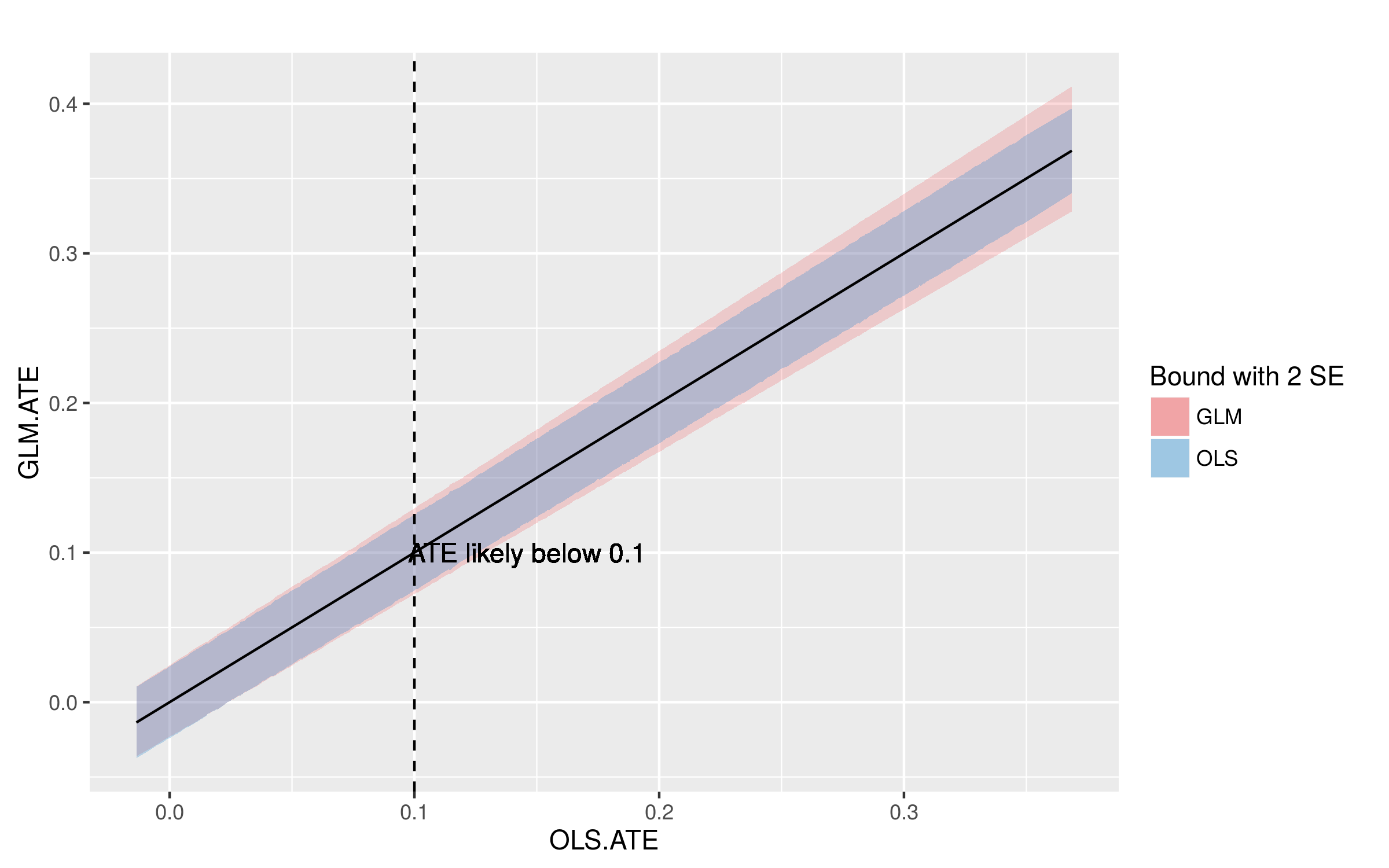}
  \caption{GLM(Poisson) vs OLS ATE estimate with 95\% Confidence Bound}
  \label{fig:ols-glm}
\end{figure}

\proglang{R}'s \texttt{lm} and \texttt{glm} functions are used to
estimated model (\ref{eq:sim-ols}) and (\ref{eq:sim-glm}). We plot the
estimated $\hat{\text{ATE}}$ from both models against each other for
each level of assumed true ATE in $\{0.001, 0.002, \dots, 0.35\}$ in
Figure \ref{fig:ols-glm}. Confidence bounds of $\hat{\text{ATE}}$ (based
on estimated standard error of $\beta$) with plus and minus $2\text{SE}$
for both models are also included.

Table \ref{tab:ols-glm} reports standard errors from both models and the
ratio of them. OLS tend to underestimate the standard error of treatment
effect. But for small ATE, this effect is not great.

\begin{table}[!htbp] \centering 
  \label{tab:ols-glm} 
\begin{tabular}{@{\extracolsep{5pt}} ccccc} 
\\[-1.8ex]\hline 
\hline \\[-1.8ex] 
 & True.ATE & OLS.se & GLM.se & Ratio \\ 
\hline \\[-1.8ex] 
1 & $0.0060$ & $0.0118$ & $0.0122$ & $0.9655$ \\ 
2 & $0.0090$ & $0.0120$ & $0.0119$ & $1.0089$ \\ 
3 & $0.0120$ & $0.0118$ & $0.0123$ & $0.9615$ \\ 
4 & $0.0150$ & $0.0119$ & $0.0123$ & $0.9709$ \\ 
5 & $0.0180$ & $0.0119$ & $0.0123$ & $0.9699$ \\ 
6 & $0.0210$ & $0.0119$ & $0.0121$ & $0.9869$ \\ 
7 & $0.0240$ & $0.0119$ & $0.0123$ & $0.9696$ \\ 
8 & $0.0270$ & $0.0122$ & $0.0125$ & $0.9789$ \\ 
9 & $0.0300$ & $0.0121$ & $0.0130$ & $0.9276$ \\ 
10 & $0.0330$ & $0.0119$ & $0.0124$ & $0.9573$ \\ 
11 & $0.0360$ & $0.0121$ & $0.0128$ & $0.9442$ \\ 
12 & $0.0390$ & $0.0122$ & $0.0127$ & $0.9592$ \\ 
\hline \\[-1.8ex] 
\end{tabular} 
\caption{GLM(Poisson) vs OLS ATE estimate and Standard Errors} 
\end{table}

\subsection{Modeling Zone Level Heterogenous
Effects}\label{modeling-zone-level-heterogenous-effects}

\label{zone}

When studying zone level (heterogenous) treatment effects, we opted to
estimate a separate model (\ref{eq:model-covariate}) subsetting our dataset
for each zone. It would have been possible to combine all zone level
treatment effects into one model and estimate on all data with the model
given in equation (\ref{eq:model-fullzone}).

\begin{equation} \label{eq:model-fullzone}
\text{clicks} \sim 1 + \text{zone} + \text{treat} + \text{treat:zone} + \text{pageNo} + \text{itemsPerPage} + \text{isLogin} + \text{catClickRateAA}
\end{equation}

The key difference between these two approaches is that model
(\ref{eq:model-covariate}) has \emph{smaller} degrees of freedom compared
to the combined models in (\ref{eq:model-fullzone}). Moreover, model
(\ref{eq:model-fullzone}) assumes the coefficients of covariates are the
same for all zones. For instance, let the coefficient for
\texttt{catClickRateAA} be denoted as $\beta_{aa}$. The single combined
regression model assumes the same $\beta_{aa}$ for all zones. In other
words, A/A stage click-through by category is assumed to be equally
powerful in explaining variances as A/B stage click-throughs for all
zones.

We argue this assumption is not realistic. For one thing, different
zones have different number of categories and very different traffic
volume. Also Table \ref{tab:reg-zone23} and \ref{tab:reg-zone46} suggest
the coefficients for $\beta_{aa}$ do in fact, vary across zones (for
zone 4, \texttt{catClickRateAA} has a larger than average explanatory
power).

Furthermore, another key assumption baked into model
(\ref{eq:model-fullzone}) is that the same residuals standard error
exists across all zones; again, Table \ref{tab:reg-zone23} and
\ref{tab:reg-zone46} provide informal evidence of violation of this
assumption.

In some studies, it is beneficial to favor a model with larger degrees
of freedom such as model (\ref{eq:model-fullzone}), when the number of
data points is small or heterogeneity in covariates and residual SEs are
not of research interest. After all, all models are unrealistic to some
extent. In our study, however, with millions of data points, we believe
the reducing degree of freedom is an acceptable tradeoff and provides
better control. In addition, we are not interested in interpreting
contributions of covariates to outcome variable \emph{across zones}, and
only care about treatment effect. Therefore, we favor model
(\ref{eq:model-covariate}) over (\ref{eq:model-fullzone}).

\subsection{Effect of Treatment
Reassignment}\label{effect-of-treatment-reassignment}

\label{reassign}

To further study the impact of treatment reassignment, we created an
indicator variable \texttt{isDoubleAssigned}. For each user, if in A/B
stage we observe she was assigned to treatment in some sessions and
control in others, we assign value 1 to \texttt{isDoubleAssigned}.
Otherwise, if a user remained in one group consistently, we assign value
0 to \texttt{isDoubleAssigned}. It's worth noting this variable is
\textbf{not} randomly assigned. It's correlated to site usage: heavy
users tend to get reassigned. For any user who visits the site more than
once with first visit and last visit spanning more than a 90 hour time,
there's a chance of getting reassigned treatment.

We included \texttt{isDoubleAssigned} as a covariate in model
(\ref{eq:model-covariate}), and results are summarized in Table
\ref{tab:double-assign}. Note the interaction term
\texttt{treat:isDoubleAssigned} is \textbf{not} included in the model.
It is tempting to estimate treatment effects based on
\texttt{isDoubleAssigned}, and consider the differences being caused by
treatment reassignment. However, since \texttt{isDoubleAssigned} is not
experimentally determined and correlates to site usage, the resulting
estimates will only be observational and not causal.

\begin{table}[!htbp] \centering 
  \caption{Effect of display of Price Range on Product Click-through} 
  \label{tab:double-assign} 
\begin{tabular}{@{\extracolsep{5pt}}lcc} 
\\[-1.8ex]\hline 
\hline \\[-1.8ex] 
 & \multicolumn{2}{c}{\textit{Dependent variable:}} \\ 
\cline{2-3} 
\\[-1.8ex] & \multicolumn{2}{c}{Product Click-through} \\ 
 & Full & Full + isDoubleAssigned \\ 
\hline \\[-1.8ex] 
 Treat & 0.0115$^{***}$ (0.0016) & 0.0115$^{***}$ (0.0015) \\ 
  Page no. & $-$0.0079$^{***}$ (0.0005) & $-$0.0079$^{***}$ (0.0005) \\ 
  Items per page & $-$0.0039$^{***}$ (0.0002) & $-$0.0039$^{***}$ (0.0002) \\ 
  Is user logged in & 0.0763$^{***}$ (0.0067) & 0.0677$^{***}$ (0.0067) \\ 
  Average click-through & 0.8700$^{***}$ (0.0090) & 0.8700$^{***}$ (0.0090) \\ 
  isDoubleAssigned &  & 0.0288$^{***}$ (0.0020) \\ 
  Constant & 0.1050$^{***}$ (0.0012) & 0.0949$^{***}$ (0.0040) \\ 
 \hline \\[-1.8ex] 
Observations & 2,438,232 & 2,438,232 \\ 
R$^{2}$ & 0.0164 & 0.0167 \\ 
Adjusted R$^{2}$ & 0.0164 & 0.0167 \\ 
Residual Std. Error & 0.7650 (df = 2438226) & 0.7650 (df = 2438225) \\ 
F Statistic & 8,124.0000$^{***}$ (df = 5; 2438226) & 6,904.0000$^{***}$ (df = 6; 2438225) \\ 
\hline 
\hline \\[-1.8ex] 
\textit{Note:}  & \multicolumn{2}{r}{$^{*}$p$<$0.1; $^{**}$p$<$0.05; $^{***}$p$<$0.01} \\ 
 & \multicolumn{2}{r}{SE are cluster-robust, clustering on per-browser-instance} \\ 
\end{tabular} 
\end{table}

The intercept term for \texttt{isDoubleAssigned} appears to be
significant - not surprisingly. This can be interpreted as heavy users
tend to have higher click through rates. In addition, standard error for
treatment effect is slightly reduced, but remains constant in its value.
\texttt{isDoubleAssigned} can be considered another covariate (which
proxies site usage), and seems useful in explaining variations in click
through. Its inclusion does not alter treatment effect estimate, because
even though \texttt{isDoubleAssigned} itself is not experimentally
determined, which group a user is assigned to still is random.

Furthermore, this comparison suggest \emph{not} to remove reassigned
users (a group that accounts for over 30\% of listing sessions) from
regression analysis. Including these users allow us to get a overall ATE
estimate applicable to the entire population, as opposed to conditional
on only light users. However, it's worth remembering the ATE estimate
will be an ``impure'' effect, compared to what the experiment initially
was designed to measure. Lastly, we choose not to include
\texttt{isDoubleAssigned} as a covariate when reporting results, even if
it reduces standard error for treatment effect. The rationale is its
interactions with other covariates and treatment is complex and not well
understood, and we would rather settle for a more conservative standard
error estimate.

\pagebreak

\section*{Reference}

\noindent [1] Belleflamme, P. (Apr, 2015). How Do Comparison Shopping
Sites Make a Living? An Update. IPDigIt,
\href{http://www.ipdigit.eu/2015/04/How-do-comparison-shopping-sites-make-a-living-an-update/}{\url{http://www.ipdigit.eu/2015/04/How-do-comparis}\ldots{}}

\hfill\break

\noindent [2] Dholakia, U. M. \& Simonson, I. (2005). The Effect of
Explicit Reference Points on Consumer Choice and Online Bidding
Behavior. Marketing Science, 24(2):206-217.
\url{http://dx.doi.org/10.1287/mksc.1040.0099}

\hfill\break

\noindent [3] Kohavi, R., Longbotham, R., Sommerfield, D., \& Henne, R.
M.. (February 2009). Controlled Experiments on the Web: Survey and
Practical Guide. Data Mining and Knowledge Discovery 18, 1 pp.~140-181.
DOI=10.1007/s10618-008-0114-1
\url{http://dx.doi.org/10.1007/s10618-008-0114-1}.

\hfill\break

\noindent [4] Niklas Schmucker. (2011). Web Tracking. SNET2 Seminar
Paper
\url{http://citeseerx.ist.psu.edu/viewdoc/download?doi=10.1.1.474.8976\&rep=rep1\&type=pdf}

\end{document}